\documentclass[twocolumn,showpacs,nofootinbib,amsmath,amssymb,prd]{revtex4}
\usepackage{hyperref}
\usepackage{mathrsfs}
\usepackage{latexsym}
\usepackage{amsmath}
\usepackage{amsbsy}
\usepackage{amssymb}
\usepackage{epsfig}
\usepackage{graphicx}
\usepackage{dcolumn}
\usepackage{graphicx}
\usepackage{bm}
\usepackage{natbib}

\newcommand{\s}{\mathcal{S}}
\begin{document}

\title{Spherically Symmetric Gravitational Collapse of General Fluids}

\author{P. D. Lasky}
	\email{paul.lasky@sci.monash.edu.au}, 

\author{A. W. C. Lun}
	\email{anthony.lun@sci.monash.edu.au}

\affiliation{Centre for Stellar and Planetary Astrophysics\\
		School of Mathematical Sciences, Monash University\\
		Wellington Rd, Melbourne 3800, Australia}
\received{December 1, 2006}
	
		\begin{abstract}
			We express Einstein's field equations for a spherically symmetric ball of general fluid such that they are conducive to an initial value problem.  We show how the equations reduce to the Vaidya spacetime in a non-null coordinate frame, simply by designating specific equations of state.  Furthermore, this reduces to the Schwarzschild spacetime when all matter variables vanish.  We then describe the formulation of an initial value problem, whereby a general fluid ball with vacuum exterior is established on an initial spacelike slice.  As the system evolves, the fluid ball collapses and emanates null radiation such that a region of Vaidya spacetime develops.  Therefore, on any subsequent spacelike slice there exists three regions;  general fluid, Vaidya and Schwarzschild, all expressed in a single coordinate patch with two free-boundaries determined by the equations.  This implies complicated matching schemes are not required at the interfaces between the regions, instead, one simply requires the matter variables tend to the appropriate equations of state.  We also show the reduction of the system of equations to the static cases, and show staticity necessarily implies zero ``heat flux''.  Furthermore, the static equations include a generalization of the Tolman-Oppenheimer-Volkoff equations for hydrostatic equilibrium to include anisotropic stresses in general coordinates.    
		\end{abstract}
		
		\maketitle
		
\section{INTRODUCTION}

Full analytic studies of gravitational collapse are intrinsically difficult due to the non-linearities in the Einstein field equations (EFEs).  Indeed, in the study of gravitational collapse, a standard approach is to approximate the region exterior to the collapse to be vacuum (see for example \cite{weinberg72,misner73} and references therein), which under the condition of spherical symmetry is the Schwarzschild spacetime.  While this simplification reduces the mathematics to tangible systems, it is physically insufficient.  The next approximation beyond this is to allow for some form of null radiation to emanate from the collapsing region \cite{lake81,lake82,bonnor89,herrera04,herrera06}.  In this way, the number of baryons are conserved within the collapsing region, however energy may still be dissipated.  In spherical symmetry, a region of outgoing null radiation is described by the Vaidya spacetime.  

The Vaidya spacetime is a solution of EFEs that contains incoherent null radiation; non-interacting particles moving along null geodesics.  References \cite{lemos92,hellaby94,gao05} have shown the Vaidya spacetime to be an extension of the dust filled Lemaitre-Tolman-Bondi (LTB) spacetime.  To do this, they proposed a LTB interior region, and at some finite radius the timelike coordinate axis tilted over to become null.  While the mathematics of this method works well, the physical realization it brings is intuitively awkward.  This is due to the exterior region of the spacetime still being in a null coordinate frame, implying the set-up is not conducive to an initial value problem. 
  
To overcome this we recast the entire problem in a non-null coordinate frame such that it can be used as an initial value problem.  In this frame, we show the incoherent null radiation of the Vaidya spacetime takes the form of a general fluid with very specific equations of state.  Therefore, rather than our interior region being dust, we formulate the interior region as a general spherically symmetric fluid such that the energy momentum terms are able to smoothly match to a Vaidya exterior.  Therefore, this article generalizes our earlier work \cite{lasky06a,lasky06b} by formulating a completely general, spherically symmetric fluid utilizing the $3+1$ formalism for general relativity which thus makes it suitable for initial value problems.  We then show the Vaidya spacetime to be a limiting case of the general fluid equations, and the Schwarzschild spacetime to be a further limiting case of this.  

Due to the method of formulation, we can prescribe data on an initial spacelike hypersurface, apply physically reasonable conditions, and the evolution of the system will take care of itself.  In particular, we discuss at length a scenario whereby the initial hypersurface is simply a ball of fluid extending out to some finite radius, and then vacuum extending beyond this to spacelike infinity.  As the system evolves, null radiation is allowed to emanate from the collapsing ball of fluid.  Thus, at any spacelike slice beyond the initial hypersurface there will be three regions of spacetime; a collapsing fluid ball, a region of outgoing null radiation and a Schwarzschild exterior extending to spacelike infinity.  Furthermore, as the Vaidya and Schwarzschild regions are shown to be limiting cases of the general fluid ball, the entire spacetime is described in terms of a single line element, expressed in a single coordinate patch.

As an expanding model is simply the time reverse of a collapsing scenario, the above model also satisfies expanding cases.  The limiting case of such collapsing or expanding models however is a static model.  As a final application we show the conditions for the set of equations derived herein to be static.  Using this we show a static spacetime necessarily has zero ``heat flux'', and furthermore show the remaining equations to generalize the Tolman-Oppenheimer-Volkoff (TOV) equations of hydrostatic equilibrium to include anisotropic stresses \cite{bowers74,poncedeleon87}.  
  
The article is set out as follows; in section \ref{IFD} we review the necessary components of the $3+1$ formalism, and derive the equations governing the general fluid system.  In section \ref{ER} we show how the equations are generalizations of both the Vaidya and Schwarzschild spacetimes, and pick out specific examples of each.  Section \ref{CP} discusses the overall picture of the gravitational collapse of a ball of fluid which emanates null radiation.  Finally, in section \ref{static} we show the reduction of the general equations to the static system.  Throughout the article we use a $+2$ signature and coordinates are denoted $x^{\mu}=\left(t,r,\theta,\phi\right)$ unless otherwise stated.  Geometrized units are used, whereby $c=G=1$.  Greek indices range from $0\ldots3$, Latin from $1\ldots3$ and all other conventions follow \cite{misner73}.

\section{General Fluid Derivation}\label{IFD}
We provide here a derivation of the general spherically symmetric fluid equations, which closely follows the method used in \cite{lasky06b}.  To begin, we define a hypersurface forming, normalized, timelike vector field, $n^{\mu}$, which has components 
\begin{align}
n^{\mu}=\frac{1}{\alpha}\left(-1,\beta,0,0\right).
\end{align}  
Here, $\alpha(t,r)>0$ is the lapse function and $\beta(t,r)$ is the radial component of the shift vector.  The energy-momentum tensor can then be irreducibly decomposed into components on and orthogonal to the spacelike hypersurface formed by the normal vector.  This decomposition is given by
\begin{align}
T_{\mu\nu}=\left(\rho +P\right)n_{\mu}n_{\nu}+Pg_{\mu\nu}+2j_{\left(\mu\right.}n_{\left.\nu\right)}+\Pi_{\mu\nu}.
\end{align}  
Here, $\rho(t,r)$ and $P(t,r)$ are respectively the energy density and isotropic pressure, 
\begin{align}
j_{\mu}=\left[\beta j(t,r),\, j(t,r),\,0,\,0\right],
\end{align}
is associated with the ``heat flux'' and $\Pi_{\mu\nu}$ is the trace-free anisotropic stress tensor.  We note here that one can further discuss each of these terms in more detail with regards to properties of the fluids [see the brief discussion regarding equation (\ref{Baryon}) below], however, the important aspect here is that this is a unique, irreducible decomposition of the energy momentum tensor.

Following \cite{lasky06b}, we define a mixed tensor according to
\begin{align}
{P_{i}}^{j}:=\text{diag}\left[-2,1,1\right].
\end{align}
Now, consider some $\left({}^{0}_{2}\right)$-tensor, $W_{\mu\nu}$ say, which is symmetric, spatial ($W_{\mu\alpha}n^{\alpha}=0$) and trace-free (${W_{\alpha}}^{\alpha}=0$).  It is trivial to show with spherical symmetry we can always write
\begin{align}
W_{ij}=w(t,r)P_{ij},
\end{align}
where $w(t,r)$ is the distinct eigenvalue of $W_{ij}$.  With spherical symmetry, many of the $3+1$ variables can be expressed in terms of their distinct eigenvalues, and we therefore make the following definitions;
\begin{itemize}
	\item The anisotropic stress tensor
		\begin{align}
		\Pi_{ij}:=\Pi(t,r)P_{ij}.
		\end{align}
	\item The trace-free extrinsic curvature
		\begin{align}
		K_{ij}-\frac{1}{3}\perp_{ij}K:=a(t,r)P_{ij}.
		\end{align}
	\item The trace-free three-Riemann tensor
		\begin{align}
		^{3}R_{ij}-\frac{1}{3}\perp_{ij}{^{3}R}:=q(t,r)P_{ij}.
		\end{align}
	\item The trace-free Hessian of the lapse function
		\begin{align}
		\frac{1}{\alpha}\left(D_{i}D_{j}-\frac{1}{3}\perp_{ij}D^{k}D_{k}\right)\alpha:=\epsilon(t,r) P_{ij}.
		\end{align}
	\item The electric curvature tensor
		\begin{align}
		E_{ij}:=C_{i\alpha j\beta}n^{\alpha}n^{\beta}:=\lambda(t,r)P_{ij}.\label{E}
		\end{align}
\end{itemize}
Here, $K_{ij}$ and $K$ are the extrinsic curvature and its trace of the spatial three-slices respectively, $\perp_{ij}$ is the three-metric on the three slices, $^{3}R_{ij}$ and $^{3}R$ are the three-Ricci tensor and scalar respectively, $D_{i}$ is the unique three-metric connection and $C_{\mu\nu\sigma\tau}$ is the Weyl conformal curvature tensor.  

As a final generalization to \cite{lasky06b}, we must include one more non-trivial metric variable in the line element.  We do this as the energy momentum tensor now has four degrees of freedom (i.e. $\rho$, $P$, $j$ and $\Pi$), implying to keep the fluid as general as possible we require a corresponding number of degrees of freedom in the metric.  Therefore, we designate the spherically symmetric line element to have the general form
\begin{align}
d\s^{2}=-\alpha^{2}dt^{2}+\frac{\left(\beta dt+dr\right)^{2}}{1+E}+R^{2}d\Omega^{2}.\label{metric}
\end{align}
Here, $E(t,r)>-1$ is an otherwise arbitrary function which reduces to the energy function in the LTB dust model \cite{lasky06a} and $d\Omega^{2}:=d\theta^{2}+\sin^{2}\theta d\phi^{2}$.  Furthermore, $R=R(t,r)$ is an arbitrary function whereby, without loss of generality, we set $R(t,0)=0$. 

The EFEs can now be expressed in $3+1$ form, and in spherical symmetry are given by two non-trivial constraint and two non-trivial evolution equations.  The Hamiltonian and momentum constraints are respectively
\begin{align}
{^{3}R}+\frac{2}{3}K^{2}-6a^{2}=&16\pi\rho,\label{ham}\\
\frac{\partial}{\partial r}\left(a+\frac{1}{3}K\right)+\frac{3a}{R}\frac{\partial R}{\partial r}=&-4\pi j,\label{mom}
\end{align}
and the two evolution equations are
\begin{align}
2\mathcal{L}_{n}K-\frac{1}{2}{^{3}R}-K^{2}-9a^{2}+\frac{2}{\alpha}D^{k}D_{k}\alpha=&24\pi P,\label{19}\\
\mathcal{L}_{n}a-aK+\epsilon-q=8\pi\Pi.\label{20}
\end{align}
Here, $\mathcal{L}_{n}$ is the Lie derivative operator with respect to the normal vector which, when acting on a scalar takes the form
\begin{align}
\mathcal{L}_{n}=\frac{1}{\alpha}\frac{\partial}{\partial t}-\frac{\beta}{\alpha}\frac{\partial}{\partial r}.
\end{align}

By putting the line element through the momentum constraint (\ref{mom}), one can show
\begin{align}
\frac{\partial R}{\partial r}\left[\frac{\mathcal{L}_{n}E}{2\left(1+E\right)}+\frac{1}{\alpha}\frac{\partial\beta}{\partial r}\right]=4\pi j R-\frac{\partial}{\partial r}\left(\mathcal{L}_{n}R\right).\label{LieE}
\end{align}

To deal with the evolution equations, we take equation (\ref{19}) and subtract six times equation (\ref{20}).  That is, we obtain an equation giving the Lie derivative of the variable $a+K/3$.  Expressing this equation in terms of the metric coefficients gives, after some algebra,
\begin{align}
1-&\left(1+E\right)\left(\frac{\partial R}{\partial r}\right)^{2}+8\pi\left(P-2\Pi\right)R^{2}\nonumber\\
=&\frac{2R\left(1+E\right)}{\alpha}\frac{\partial\alpha}{\partial r}\frac{\partial R}{\partial r}-2R{\mathcal{L}_{n}}^{2}R-\left(\mathcal{L}_{n}R\right)^{2},\label{other}
\end{align}
where ${\mathcal{L}_{n}}^{2}R:=\mathcal{L}_{n}\left(\mathcal{L}_{n}R\right)$.  It is both interesting and pertinent to note that the two evolution equations, (\ref{19}) and (\ref{20}), are dependant on one another due to the spherical symmetry.  That is, these two equations can be reduced, without loss of generality, to the single evolution equation given by (\ref{other}).  Substituting this through the Hamiltonian constraint (\ref{ham}), and utilizing the momentum constraint (\ref{mom}), gives the following equation,
\begin{align}
\frac{\partial}{\partial r}&\left[\frac{R^{2}\left(1+E\right)}{\alpha}\frac{\partial\alpha}{\partial r}\frac{\partial R}{\partial r}-R^{2}{\mathcal{L}_{n}}^{2}R-4\pi\left(P-2\Pi\right)R^{3}\right]\nonumber\\
&=4\pi\left(\rho\frac{\partial R}{\partial r}+j\mathcal{L}_{n}R\right)R^{2}.\label{this}
\end{align} 
We now define a function, $M(t,r)$, according to 
\begin{align}
\frac{\partial M}{\partial r}:=4\pi\left(\rho\frac{\partial R}{\partial r}+j\mathcal{L}_{n}R\right)R^{2}.\label{massdef}
\end{align}
Regularity of $\rho$ and $j$ at $r=0$, and given $R(t,0)=0$ implies $M(t,0)=0$.  We refer to this function as a {\it mass} function as in the dust case it reduces to the mass function of the LTB metric \cite{lasky06a}, and to the Schwarzschild mass in the vacuum case.  Furthermore, we will show as one of our main results that it also reduces to the mass function of the Vaidya system (see section \ref{Vadsec}).  

Now, integrating both sides of equation (\ref{this}) gives
\begin{align}
\frac{M}{R^{2}}+4\pi\left(P-2\Pi\right)R=\frac{1+E}{\alpha}\frac{\partial\alpha}{\partial r}\frac{\partial R}{\partial r}-{\mathcal{L}_{n}}^{2}R,\label{that}
\end{align}
where the specification of $R(t,0)=0$ and regularity at $r=0$ ensures any functions of integration vanish.

Now, putting equation (\ref{that}) back through (\ref{other}), we find an equation for the square of $\mathcal{L}_{n}R$, which is
\begin{align}
\left(\mathcal{L}_{n}R\right)^{2}=\frac{2M}{R}+\left(1+E\right)\left(\frac{\partial R}{\partial r}\right)^{2}-1.\label{LieR}
\end{align}
We now substitute this into equation (\ref{that}), and also utilize equation (\ref{LieE}) to find the Lie derivative for the mass function
\begin{align}
\mathcal{L}_{n}M=-4\pi R^{2}\left[\left(P-2\Pi\right)\mathcal{L}_{n}R+j\frac{\partial R}{\partial r}\left(1+E\right)\right].\label{LieM}
\end{align}

The conservation of energy-momentum equations, $\nabla_{\alpha}{T^{\alpha}}_{\mu}=0$, provide the remaining pieces of information regarding the spacetime.  When decomposed onto and orthogonal to the spacelike hypersurfaces these become
\begin{align}
0=&\left(\mathcal{L}_{n}-K\right)\rho+D^{k}j_{k}+2\dot{n}^{k}j_{k}-PK-6\Pi a,\\
0=&\left(\mathcal{L}_{n}-K\right)j_{i}+\left(D^{k}+\dot{n}^{k}\right)\left(\Pi_{ik}+\perp_{ik}P\right)+\rho\dot{n}_{i},
\end{align}
where $\dot{n}_{i}$ is the four-acceleration.  By putting the line element through these equations, they respectively become 
\begin{align}
&\mathcal{L}_{n}\rho=\left(\rho+P-2\Pi\right)\left[\frac{1}{2\left(1+E\right)}\mathcal{L}_{n}E+\frac{1}{\alpha}\frac{\partial\beta}{\partial r}-\frac{2}{R}\mathcal{L}_{n}R\right]\nonumber\\
&-j\left(1+E\right)\left[\frac{1}{j}\frac{\partial j}{\partial r}+\frac{1}{2\left(1+E\right)}\frac{\partial E}{\partial r}+\frac{2}{R}\frac{\partial R}{\partial r}+\frac{2}{\alpha}\frac{\partial\alpha}{\partial r}\right]\nonumber\\
&-\frac{6\Pi}{R}\mathcal{L}_{n}R,\label{cont}\\
&\frac{\mathcal{L}_{n}\left(j\sqrt{1+E}\right)}{\sqrt{1+E}}=2j\left[\frac{1}{2\left(1+E\right)}\mathcal{L}_{n}E+\frac{1}{\alpha}\frac{\partial\beta}{\partial r}-\frac{1}{R}\mathcal{L}_{n}R\right]\nonumber\\
&-\frac{\left(\rho+P-2\Pi\right)}{\alpha}\frac{\partial\alpha}{\partial r}-\frac{\partial}{\partial r}\left(P-2\Pi\right)+\frac{6\Pi}{R}\frac{\partial R}{\partial r}.\label{Euler}
\end{align}
The two conservation equations expressed above provide evolution equations for two of the four matter variables, namely $\rho$ and $j$.  Therefore, the only way of evolving the other two matter variables, $P$ and $\Pi$, is to employ two equations of state to supplement the system.  Furthermore, we note the system of equations is currently underdetermined.  One simple way of expressing this is to note the differences between the general fluid system and the more restrictive perfect fluid system presented in \cite{lasky06b}, which was fully determined.  The key differences between the two systems are that we now have two more matter variables, $j$ and $\Pi$, and one more geometric variables, $R$, however, only two extra equations have been introduced, which are (\ref{cont})\footnote{In the perfect fluid case, this equation is an identity which is the derivative of equation (\ref{LieM}).} and one more equation of state.    

The system being underdetermined implies that to find solutions for various physical scenarios requires the input of extra physical or geometric effects.  One possible way is to introduce an equation governing the conservation of Baryon number.  In this way, Baryon number will be conserved for the interior fluid region, implying only non-baryonic energy may escape the collapse region.  This conservation equation is usually expressed as (see for example \cite{ehlers93,font03})
\begin{align}
\nabla_{\alpha}\left(\rho_{0}u^{\alpha}\right)=0,\label{Baryon}
\end{align} 
where $\rho_{0}$ is the rest-mass energy density, and $u^{\mu}$ is the comoving four-vector.  The introduction of this equation will close the system, however its introduction requires the additional consideration and discussion of thermodynamic quantities, a task reserved for future work.

Summarily, the system of equations describing a general spherically symmetric fluid are the line element (\ref{metric}) and the equations governing the metric coefficients (\ref{LieE}), (\ref{LieR}) and (\ref{LieM}), along with two equations of state and the two conservation equations, (\ref{cont}) and (\ref{Euler}).

\subsection{Tidal Forces}
It is an interesting exercise to calculate the tidal forces for the general fluid spacetime, and thereafter for the resulting subclasses of Vaidya and Schwarzschild spacetimes.  The tidal forces are given by the electric curvature tensor \cite{ellis71}, $E_{ij}$, which is given by the contraction of two indices of the Weyl tensor with respect to the normal vector [see equation (\ref{E})].  A constraint equation from the gravito-electromagnetic system of equations, when expressed in terms of metric coefficients, is\footnote{The magnetic component of the Weyl tensor vanishes identically in spherical symmetry.}
\begin{align}
\frac{\partial}{\partial r}&\left(\lambda+4\pi\Pi+\frac{4\pi}{3}\rho\right)+\frac{3}{R}\frac{\partial R}{\partial r}\left(\lambda+4\pi\Pi\right)=\frac{4\pi j}{R}\mathcal{L}_{n}R.
\end{align}
Substituting equation (\ref{massdef}) into the above,
we find an equation which can be integrated to give
\begin{align}
\lambda=\frac{M}{R^{3}}-\frac{4\pi}{3}\left(\rho+3\Pi\right).\label{tidal}
\end{align}
Here, we have used $R(t,0)=0$ and regularity of $\lambda$, $\rho$ and $\Pi$ at $r=0$.  We will see in the following sections how this becomes even simpler for both the Vaidya and Schwarzschild spacetimes.

\section{Exterior regions}\label{ER}
Having derived the fluid equations we can show how, by specifying particular equations of state, these reduce to both the Schwarzschild and Vaidya spacetimes.  This implies we can establish collapse scenarios by having an interior fluid, and an exterior region being either Schwarzschild or Vaidya.  Moreover, the entire spacetime will be described in a single coordinate system, by simply taking the interior metric and letting the matter variables tend to the appropriate equation of state at some finite radius.       
\subsection{Schwarzschild}\label{Schwsec}
The reduction to the Schwarzschild spacetime is simply given by having all the energy-momentum variables vanish at some radius (i.e. $\rho=P=j=\Pi=0$).  Equation (\ref{massdef}) and (\ref{LieM}) then imply that the mass function is a constant, i.e. $M=M_{s}$.  The tidal forces for the spacetime are now given by
\begin{align}
\lambda=\frac{M_{s}}{R^{3}}.
\end{align}  
This is a slightly more general result than the standard form (see for example \cite{lasky06a}) due to the function $R$ determining the sizes of the two-spheres.

There now remains four metric coefficients, $\alpha$, $\beta$, $E$ and $R$, and just two non-trivial equations, which are
\begin{align}
\frac{\partial R}{\partial r}\left[\frac{\mathcal{L}_{n}E}{2\left(1+E\right)}+\frac{1}{\alpha}\frac{\partial\beta}{\partial r}\right]=-\frac{\partial}{\partial r}\left(\mathcal{L}_{n}R\right),\label{ko}\\
\left(\mathcal{L}_{n}R\right)^{2}=\frac{2M_{s}}{R}+\left(1+E\right)\left(\frac{\partial R}{\partial r}\right)^{2}-1.\label{k1}
\end{align}
Therefore, the line element (\ref{metric}) where $\alpha$, $\beta$, $E$ and $R$ satisfy equations (\ref{ko}) and (\ref{k1}) describe an infinite family of coordinates for the Schwarzschild spacetime.  This system is underdetermined as there are two differential equations for four variables, and as such, one may make any suitable choice for any of the four metric coefficients, provided equations (\ref{ko}) and (\ref{k1}) can be satisfied.  

Whilst this is a perversely general system describing the relatively simple Schwazrschild spacetime, we should note that the amount of generality is a necessary evil.  This is because, in order to describe the entire system in a single coordinate patch as per our initial aim, one must first establish the interior, matter filled region, and then use the coordinate freedom in the exterior to express the Schwarzschild region in appropriate coordinates.  Thus, the amount of freedom in the exterior corresponds to the vast number of solutions for the general fluid equations for the interior.  As an example, if we consider matching a general fluid to a vacuum region (with no intermediate Vaidya region) then, as in the case of \cite{lasky06b}, we demand there exist a small region of dust between the fluid and the vacuum.  This dust region has all matter terms vanish with the exception of $\rho$.  Equation (\ref{Euler}) then implies the lapse is only a function of the temporal coordinate, and hence coordinate freedom can be utilized to set it to unity without loss of generality.  Therefore, to match the vacuum region to the dust region, we also set $\alpha=1$ in the Schwarzschild region.   

With $\alpha=1$ for the Schwarzschild region, many familiar solutions arise.  This coordinate choice corresponds to a vanishing four-acceleration, $\dot{n}^{i}=0$, and thus the normal vector is tangent to a congruence of timelike geodesics.  If we make the further choice $R=r$, then the system reduces to the familiar generalized Painleve-Gullstrand (GPG) coordinates \cite{lasky06a,lasky06b}.  Another possibility is to set $\beta=0$, which implies equation (\ref{ko}) can be integrated 
and we find the Novikov coordinates
\begin{align}
d\s^{2}=-dt^{2}+\frac{1}{f}\left(\frac{\partial R}{\partial r}\right)^{2}dr^{2}+R^{2}d\Omega^{2},\label{nov1}
\end{align}
where $f=f(r)$ and $R$ is a solution of
\begin{align}
\left(\frac{\partial R}{\partial t}\right)^{2}=\frac{2M_{s}}{R}+f-1.\label{nov2}
\end{align}

We have shown how a fluid matches directly to a Schwarzschild region.  However, our aim was to include radiation into the system, which implies we will actually be matching the fluid to a Vaidya region.  Furthermore, the Vaidya region will match onto a Schwarzschild region along a null line.  To establish these conditions, we must first establish the reduction of the general fluid equations to the Vaidya system.



\subsection{Vaidya}\label{Vadsec}
It is not intuitively obvious how one retrieves the Vaidya spacetime from the system of equations derived herein.  This is because we do not know {\it a priori} appropriate equations of state for the Vaidya spacetime when written using non-null coordinates.  For this reason we derive the results of this section in reverse.  That is, we start with the Vaidya line element in standard coordinates, $\left(u,R,\theta,\phi\right)$, which is given by
\begin{align}
d\s^{2}=-\left(1-\frac{2m}{R}\right)du^{2}-2dudR+R^{2}d\Omega^{2},\label{vadmetric}
\end{align}
where $u$ is an outgoing null coordinate and $m(u)$ is the Vaidya mass function satisfying 
\begin{align}
\frac{dm}{du}=4\pi\rho R^{2}.\label{vadmass}
\end{align}
We have denoted $R$ to be the radial coordinate of the Vaidya metric such that we can perform a coordinate transformation given by $R=R(t,r)$ and $u=u(t,r)$, where $t$ and $r$ are the temporal and radial coordinates of the general fluid line element respectively.  We define the simultaneous coordinate transformation to be
\begin{align}
\frac{\partial R}{\partial t}=-\frac{\alpha m}{R}-\frac{\beta\left(1-m/R\right)}{\sqrt{1+E}},&\,\,\,\,\frac{\partial R}{\partial r}=\frac{-\left(1-m/R\right)}{\sqrt{1+E}},\nonumber\\
\frac{\partial u}{\partial t}=-\alpha+\frac{\beta}{\sqrt{1+E}},&\,\,\,\,\frac{\partial u}{\partial r}=\frac{1}{\sqrt{1+E}},\label{Rs}
\end{align}
which renders the line element in the exact form of equation (\ref{metric}).  The above equations imply the important relation
\begin{align}
\mathcal{L}_{n}R=-\frac{m}{R}.\label{LieRvad}
\end{align}  
Furthermore, the integrability conditions must be satisfied for this to be a valid coordinate transformation.  These integrability conditions provide restrictions for the metric coefficients which, after some algebra, are found to be 
\begin{align}
\frac{\sqrt{1+E}}{\alpha}\frac{\partial\alpha}{\partial r}&=\frac{\mathcal{L}_{n}E}{2\left(1+E\right)}+\frac{1}{\alpha}\frac{\partial\beta}{\partial r}\nonumber\\
&=\frac{-1}{R}\left(\mathcal{L}_{n}m+\sqrt{1+E}\frac{\partial m}{\partial r}\right)-\frac{m}{R^{2}}.\label{theseones}
\end{align} 
Therefore, the Vaidya spacetime is given by the line element (\ref{metric}), whereby the metric coefficients satisfy equations (\ref{Rs}-\ref{theseones}).  It is straightforward to see this system is underdetermined and therefore, as with the Schwarzschild case, there are infinitely many ways of expressing the Vaidya spacetime in these general spherically symmetric coordinates.  That is, any solution of the above equations represents a Vaidya spacetime, the choice of which depends on the physical problem being studied.    

We can now put these conditions through the reduced field equations we have derived to obtain the equation of state for the Vaidya spacetime when expressed in non-null coordinates.  Beginning with equation (\ref{LieR}) we find the remarkable result that
\begin{align}
m=M.\label{mism}
\end{align}  
That is, the mass we derived with the definition given by equation (\ref{massdef}) reduces to the Vaidya mass in the appropriate limit.  This function was also shown in \cite{lasky06b} to reduce to the Schwarzschild and LTB masses in the appropriate limits.  

Now, if we put the integrability conditions through equations (\ref{LieE}) and (\ref{that}), after some algebra we can show
\begin{align}
\mathcal{L}_{n}M+\sqrt{1+E}\frac{\partial M}{\partial r}=4\pi R^{2}\left(P-2\Pi-j\sqrt{1+E}\right).\label{one}
\end{align}
Furthermore, putting the integrability conditions through equations (\ref{massdef}) and (\ref{LieM}), we find respectively
\begin{align}
\frac{\partial M}{\partial r}=&\frac{-4\pi R^{2}}{\sqrt{1+E}}\left[\rho-\frac{M}{R}\left(\rho-j\sqrt{1+E}\right)\right],\label{two}\\
\mathcal{L}_{n}M=&4\pi R^{2}\left[\frac{M}{R}\left(P-2\Pi-j\sqrt{1+E}\right)+j\sqrt{1+E}\right].\label{three}
\end{align}
Finally, combining equations (\ref{one}-\ref{three}), we derive an algebraic relation between all energy momentum variables
\begin{align}
\rho+P-2\Pi-2j\sqrt{1+E}=0.\label{schmoo}
\end{align}
Now, putting the above information into the conservation equations (\ref{cont},\ref{Euler}), and combining these we find
\begin{align}
\left(\mathcal{L}_{n}+\sqrt{1+E}\frac{\partial}{\partial r}\right)&\left(\rho-j\sqrt{1+E}\right)\nonumber\\
&=\frac{2}{R}\left(j\sqrt{1+E}+3\Pi\right).
\end{align}
The remaining piece of information to be utilized is the form of the mass in the Vaidya solution, i.e. equation (\ref{vadmass}).  By inverting the coordinate transformation and expanding the derivative with respect to the null coordinate in terms of radial and temporal derivatives, one can show
\begin{align}
\frac{dm}{du}=4\pi R^{2}j\sqrt{1+E},
\end{align}
where we have also made use of equations (\ref{one}-\ref{schmoo}).  Hence, direct comparison with equation (\ref{vadmass}) implies $\rho=j\sqrt{1+E}$.  Finally, combining all information, we find the equations of state governing the Vaidya spacetime are
\begin{align}
\rho=3P=j\sqrt{1+E}=-3\Pi.\label{vadEOS}
\end{align}
It is interesting that $\rho=3P$ is a radiative equation of state, and hence this aspect of the result is one that is expected.

We can put this equation of state through equation (\ref{tidal}) to show the tidal forces for the Vaidya region of the spacetime are given by the simple result
\begin{align}
\lambda=\frac{M}{R^{3}}.
\end{align}
This is the same form as for the Schwarzschild spacetime, however we note the mass in the Schwarzschild is a constant, whereas it is still a function of both the temporal and radial coordinates in the Vaidya case. 

There are two further results from this section that will prove fruitful.  Firstly, putting the equations of state through equation (\ref{one}) we find 
\begin{align}
\mathcal{L}_{n}M+\sqrt{1+E}\frac{\partial M}{\partial r}=0.\label{thy}
\end{align}
Also putting the equations of state back through equation (\ref{theseones}), we find the lapse function is highly constrained
\begin{align}
\frac{1}{\alpha}\frac{\partial\alpha}{\partial r}=\frac{-M}{R^{2}\sqrt{1+E}}.\label{will}
\end{align}
This equation has some interesting physics in that the right hand side only vanishes if the mass function vanishes.  Therefore, providing $M\neq0$, the lapse function must necessarily be a function of the radial coordinate, and hence we are never free to set the lapse function to unity for the Vaidya spacetime.  Another way to state this result is that in the Vaidya spacetime, the unit timelike vector normal to spacelike slices, $n^{\mu}$, can not be tangent to timelike geodesics.    

Furthermore, this equation gives a matching condition for the Vaidya and Schwarzschild regions.  As mentioned, the Schwarzschild system is vastly underdetermined, however, when matching Schwarzschild to Vaidya we require the lapse be a solution of (\ref{will}) in the vacuum region (with $M=M_{s}$) such that it matches continuously across the interface.  This point is discussed in more detail in section \ref{matcheg}.

As mentioned, the Vaidya system of equations is underdetermined.  It is interesting to note that the choice of functions $R=r$ is not a valid one for the Vaidya spacetime.  This is because choosing $R=r$ reduces the system to two differential equations, the general solution of which has the mass function being a constant.  Hence, choosing $R=r$ forces the Vaidya system to give the Schwarzschild solution.  Therefore, establishing a radiating problem necessitates the extra degree of freedom we put into the line element in the form of the general function $R$.    

\section{Complete Picture}\label{CP}
\subsection{Summary of Equations}
We provide here a summary of the important equations for the general fluid system, as well as for the reduction to both the Vaidya and Schwarzschild spacetimes.  This enables us to see the appropriate limits one needs to take at the boundaries between any two regions. 
\subsubsection{General Fluid}
There are three main evolution equations resulting from EFEs.  These, and the mass equation, are given by
\begin{subequations}\label{IF}
\begin{align}
\frac{R^{\prime}\mathcal{L}_{n}E}{2\left(1+E\right)}=&4\pi jR-\frac{R^{\prime}}{\alpha}\frac{\partial\beta}{\partial r}-\frac{\partial}{\partial r}\left(\mathcal{L}_{n}R\right),\\
\left(\mathcal{L}_{n}R\right)^{2}=&\frac{2M}{R}+\left(1+E\right){R^{\prime}}^{2}-1,\\
\mathcal{L}_{n}M=&-4\pi R^{2}\left[\left(P-2\Pi\right)\mathcal{L}_{n}R+jR^{\prime}\left(1+E\right)\right],\\
\frac{\partial M}{\partial r}=&4\pi\left(\rho R^{\prime}+j\mathcal{L}_{n}R\right)R^{2},\label{IFMass}
\end{align}
\end{subequations}
where $R^{\prime}:=\partial R/\partial r$.  The two conservation equations, (\ref{cont}) and (\ref{Euler}), are also required for the system.

\subsubsection{Vaidya}
The Vaidya system is given by the equation of state (\ref{vadEOS}), and the same three evolution equations and mass definition become 
\begin{subequations}\label{Vad}
\begin{align}
\frac{R^{\prime}\mathcal{L}_{n}E}{2\left(1+E\right)}=&\frac{4\pi\rho R}{\sqrt{1+E}}-\frac{R^{\prime}}{\alpha}\frac{\partial\beta}{\partial r}-\frac{\partial}{\partial r}\left(\mathcal{L}_{n}R\right),\\
\left(\mathcal{L}_{n}R\right)^{2}=&\frac{2M}{R}+\left(1+E\right){R^{\prime}}^{2}-1,\label{VadLieE}\\
\mathcal{L}_{n}M=&4\pi\rho R^{2},\\
\frac{\partial M}{\partial r}=&\frac{-4\pi\rho R^{2}}{\sqrt{1+E}}.
\end{align}
Now, the two conservation equations have reduced to a single equation,
\begin{align}
\mathcal{L}_{n}\rho+\sqrt{1+E}\frac{\partial\rho}{\partial r}=&\frac{2\rho}{R}
\end{align}
\end{subequations}

\subsubsection{Schwarzschild}
For the Schwarzschild region, all the matter terms vanish, and hence the evolution equations are
\begin{subequations}\label{Schw}
\begin{align}
\frac{R^{\prime}\mathcal{L}_{n}E}{2\left(1+E\right)}=&-\frac{R^{\prime}}{\alpha}\frac{\partial\beta}{\partial r}-\frac{\partial}{\partial r}\left(\mathcal{L}_{n}R\right),\label{schw1}\\
\left(\mathcal{L}_{n}R\right)^{2}=&\frac{2M}{R}+\left(1+E\right){R^{\prime}}^{2}-1,\\
\mathcal{L}_{n}M=&0,\label{trivial}\\
\frac{\partial M}{\partial r}=&0,\end{align}
\end{subequations}
where the conservation equations are now trivially satisfied.  

\subsection{Generic Collapse Model}
\begin{figure}[h]
		\begin{center}
		\includegraphics[height=0.3\textwidth,width=0.3\textwidth]{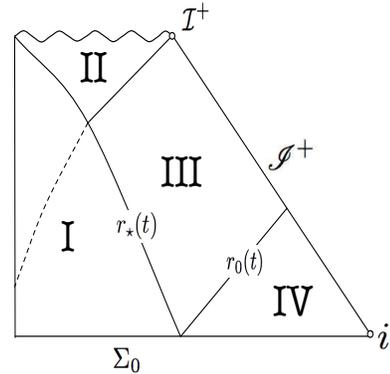}\caption{\label{diag} Compactification diagram for the collapse of a ball of fluid: $i$ represents spacelike infinity, $\mathscr{I}^{+}$ is future null infinity, $\mathcal{I}^{+}$ is future timelike infinity and $\Sigma_{0}$ is the initial spacelike hypersurface.}  
		\end{center}
\end{figure}

While analytic solutions of the general fluid equations (\ref{IF}) are obviously difficult to find, we can still analyse particular aspects of the entire system [including equations (\ref{Vad}) and (\ref{Schw})].  In particular, we consider a scenario whereby we define the data on an initial spacelike hypersurface (denoted $\Sigma_{0}$ in figure \ref{diag}) such that a ball of spherically symmetric fluid reaches out to a finite radius of $r=r_{\star}(t=0)$, such that vacuum exists for $r>r_{\star}(0)$.  

We then allow the system to evolve, with the condition that the interior fluid ball emanates only null radiation.  At present this condition is put in manually by asserting the boundary of the fluid region to be where the equations of state for the interior reduce to the equation of state for the Vaidya system, i.e. equation (\ref{vadEOS}).  We therefore have the boundary, $r_{\star}(t)$, separating the region of general fluid, region I of figure \ref{diag}, and the Vaidya region, denoted region III.  This Vaidya region will have begun on the initial hypersurface with zero volume, and be expanding along the null cone in the direction of increasing radius, extending out to future null infinity, denoted $\mathscr{I}^{+}$.  Exterior to this Vaidya region is a region of Schwarzschild spacetime, denoted as region IV and governed by equations (\ref{Schw}), extending out to spacelike infinity, $i$.    

We note the other condition we have imposed for this scenario is that we used an equation of state and initial conditions such that the radius of the fluid region reduces to zero volume in a finite time.  Using equation (\ref{IF}-\ref{Schw}), we can look at what happens when this collapse point is reached.  In particular, when $r_{\star}\rightarrow0$, we note the energy density will diverge.  The mass function will then become a constant, on the proviso that the $j$ does not diverge.  The mass function becoming a constant is enough to imply the system of general fluid equations (\ref{IF}) reduce to the system of Schwarzschild equations (\ref{Schw}).  Therefore, as soon as $r_{\star}=0$ a Schwarzschild black hole remains, as indicated by figure \ref{diag}.  

We further look at the apparent horizon, which is indicated schematically by the dashed line in figure \ref{diag}.  This will evolve from $r=0$ at a finite time, and eventually intersect with the surface of the fluid ball, $r=r_{\star}$.  The interior fluid region will continue from this point unaffected, however, the exterior will no longer be the Vaidya system.  This is due to the surface, $r_{\star}$, being inside the apparent horizon, implying the null cones have tilted over to such an extent that null radiation can no longer escape.  Therefore, the region interior of the event horizon, but still exterior to the surface $r=r_{\star}$ is Schwarzschild (region II).  This event horizon extends out to future timelike infinity, $\mathcal{I}^{+}$.

\begin{figure}[h]
		\begin{center}
		\includegraphics[height=0.3\textwidth,width=0.3\textwidth]{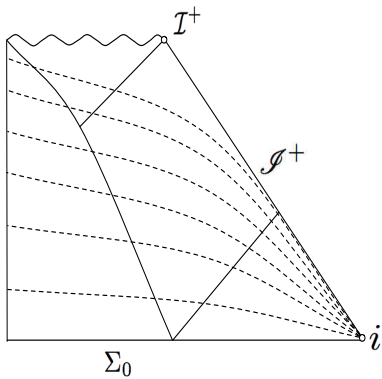}\caption{\label{diag2} Compactification diagram showing the evolution of spacelike hypersurfaces}
		\end{center}
\end{figure}

As a final aid, figure \ref{diag2} shows a schematic of spacelike slices through the compactification diagram.  We note that each slice, schematically represented by the dashed lines, begins in the general fluid region, extends through Vaidya, into the Schwarzschild region and out to spacelike infinity, $i$.  

\subsection{Matching example}\label{matcheg}
Specific solutions of the model we have established will usually be found by first constructing the fluid region, and using this to determine the form of the regions exterior to it, i.e. the Vaidya and Schwarzschild regions.  However, analytic solutions for the interior are difficult to come by and will probably require numerical analysis.  Therefore, as an illustrative example, we find here a particular example for the Vaidya region, and show how this reduces to the Schwarzschild region in the same coordinates.  In particular, the example we choose is the marginally bound case, such that $E=0$.  This is the reverse scenario to how one would tackle a physical problem, and we shall see that this implies the Vaidya region is underdetermined, as it requires the input of boundary conditions coming from the fluid region.    

The amount of coordinate freedom in the vacuum region implies we can stipulate conditions on the metric functions according to the form of these functions in the Vaidya region.  That is, in the Vaidya region, equations (\ref{Rs}) and (\ref{will}) (now with $E=0$) hold, so without loss of generality we can stipulate these to hold for the Schwarzschild region also.  This will imply both regions will be expressed in a single coordinate patch, as per our initial aim.  In the Schwarzschild region the mass becomes a constant, and equations (\ref{Rs}) and (\ref{will}) become
\begin{align}
\frac{\partial R}{\partial r}=&-\left(1-\frac{M_{s}}{R}\right),\label{Rreg}\\
\frac{\partial R}{\partial t}=&-\beta\left(1-\frac{M_{s}}{R}\right)-\frac{\alpha M_{s}}{R},\label{Rteg}\\
\frac{1}{\alpha}\frac{\partial\alpha}{\partial r}=&-\frac{M_{s}}{R^{2}}.\label{aleg}
\end{align}
Furthermore, simplifying equation (\ref{schw1}) with $E=0$ and the above equations gives
\begin{align}
\frac{1}{\alpha}\frac{\partial\beta}{\partial r}=-\frac{M_{s}}{R^{2}}.\label{beeg}
\end{align}
These four equations can be solved to give an exact solution for the exterior Schwarzschild region when matched on to a marginally bound Vaidya region.  Firstly, to integrate equation (\ref{Rreg}) we require the {\it Lambert W} function, $W(\xi)$, which is defined such that it is a solution of the algebraic equation (see for example \cite{weisstein98})
\begin{align}
W(\xi)\exp\left[W(\xi)\right]=\xi.
\end{align}
Then, by defining
\begin{align}
\xi:=\frac{-1}{M_{s}}\exp\left\{\frac{-1}{M_{s}}\left[M_{s}+r+f(t)\right]\right\},
\end{align}
where $f(t)$ is an arbitrary function of integration, we find the solution of equation (\ref{Rreg}) is
\begin{align}
R=M_{s}\left[W(\xi)+1\right].\label{Rvadsoln}
\end{align}
The function $f(t)$ is given by utilizing a more specific boundary condition between the Vaidya and Schwarzschild regions.  It is interesting to note at the end point of collapse (when the entire solution is simply given by the Schwarzschild spacetime), we can use the boundary condition that $R(t,0)=0$, which implies $f(t)=-M_{s}\ln M_{s}$, which further simplifies the form of $\xi$. 

Now, putting (\ref{Rvadsoln}) through the remaining three equations, we find 
\begin{align}
\alpha=&\frac{W(\xi)}{W(\xi)+1},\label{alvadsoln}\\
\beta=&\frac{-1}{W(\xi)+1}+\frac{df}{dt},\label{bevadsoln}
\end{align}
where we have used coordinate freedom to re-scale an arbitrary function of $t$ to unity in the solution of the lapse function.  Thus, the line element (\ref{metric}), where the metric coefficients are given by equations (\ref{Rvadsoln}-\ref{bevadsoln}) with $E=0$, describes the Schwarzschild spacetime in a coordinate system which matches continuously to a marginally bound, non-null coordinate system for the Vaidya spacetime.  

While we have solved for the exterior Schwarzschild region, this does not provide enough information to solve for the Vaidya region.  We can show in the Vaidya region that equations (\ref{will}) and (\ref{VadLieE}) imply
\begin{align}
\alpha=\beta-\frac{df}{dt},\label{be}
\end{align}
which further implies equation (\ref{thy}) can be integrated to give
\begin{align}
M=M\left[r+f(t)\right].
\end{align}
If we integrate equations (\ref{Rs}) for the null coordinate $u$, we find $u=r+f(t)$, which implies the expected result that the mass is a function of the null coordinate, $M=M(u)$.  Therefore, characteristic curves of the mass function are the null lines in the spacetime.  Moreover, this implies that the boundary between the Vaidya and Schwarzschild region, $r_{0}(t)$, is given by a specific characteristic, implying this boundary is null.

In order to integrate the remainder of the system entirely, we need to solve
\begin{align}
\frac{\partial R}{\partial r}=-\left(1-\frac{M}{R}\right),\label{hyy}
\end{align}  
which is an Abel equation of the second kind.  However, to integrate it we require more knowledge of the specific form of the function $M$, which is to be gained from boundary conditions.  Unfortunately, the boundary at $r_{0}(t)$ offers no new information on the form of the mass function.  Intuitively, this is due to the causality of the system, which essentially works from the interior to the exterior regions (i.e. the interior regions influence the exterior regions).  Therefore, the boundary conditions required to solve the remainder of the system are given by the boundary at $r_{\star}$.  Hence, once a specific fluid region is found, it can be used to stipulate boundary conditions at $r_{\star}$, which can then be used to fully solve equation (\ref{hyy}).  Once this equation is solved, equation (\ref{will}) can be integrated to find the lapse function, equation (\ref{be}) implies we know $\beta$, and we can use the remaining equation on $\partial R/\partial t$ to relate the arbitrary functions.  Therefore, the knowledge of the functional form of $M$, given by the boundary conditions at $r_{\star}$, will imply the Vaidya region is completely soluble.  

\section{static solution}\label{static}
\begin{figure}[ht]
		\begin{center}
		\includegraphics[height=0.3\textwidth,width=0.3\textwidth]{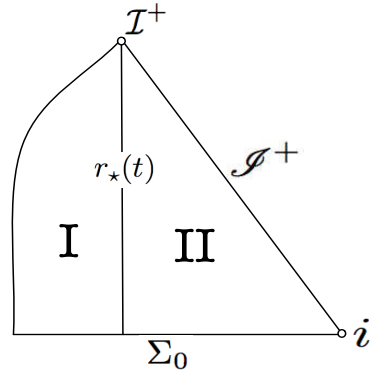}\caption{\label{diag3} Compactification diagram for static scenarios}
		\end{center}
\end{figure}

There is the situation whereby a model is not collapsing or expanding, but is static.  In reality, while a star may appear {\it static}, it will still radiate, and hence is not {\it static} in the relativistic sense.  A purely static system however will have zero fluxes, implying it will not radiate and hence the exterior spacetime is simply Schwarzschild.  The compactification diagram for this scenario is given in figure \ref{diag3}, and this section is therefore devoted to using our derivation of the general fluid equations to find conditions for the spacetime to be static.  In the process we generalize the TOV equations of hydrostatic equilibrium.  We note however that there should further exist solutions to our equations that are not static, but the boundary of the fluid region remains constant.  These solutions will be sought in a future article.    

A static spacetime is defined such that there exists a timelike, Killing vector field which is everywhere hypersurface orthogonal \cite{wald84}.  The normal vector we have used is both timelike and hypersurface orthogonal, however it being normalized places too tight a restriction on the function.  We therefore generalize the normal vector by multiplying by an arbitrary function, such that 
\begin{align}
N^{\mu}:=\mathcal{F}n^{\mu}, 
\end{align}
where $\mathcal{F}=\mathcal{F}(t,r)$.  It is trivial to show $N^{\mu}$ is also timelike and hypersurface forming.  Therefore, solving Killings equations with respect to this vector field will provide conditions for staticity of the spacetime.

Killings equations with respect to $N^{\mu}$ are
\begin{align}
N^{\alpha}\frac{\partial g_{\mu\nu}}{\partial x^{\alpha}}+g_{\mu\alpha}\frac{\partial N^{\alpha}}{\partial x^{\nu}}+g_{\alpha\nu}\frac{\partial N^{\alpha}}{\partial x^{\mu}}=0,
\end{align}
which provides a system of ten coupled differential equations.  While a majority are trivially satisfied, we find evaluating the $\left(\theta,\theta\right)$ or $\left(\phi,\phi\right)$ components gives
\begin{align}
\mathcal{L}_{n}R=0.\label{con1}
\end{align}
Likewise, the $\left(r,r\right)$ component implies
\begin{align}
\frac{\mathcal{L}_{n}E}{2\left(1+E\right)}+\frac{1}{\alpha}\frac{\partial\beta}{\partial r}=0.\label{con2}
\end{align}
Evaluating the $\left(t,r\right)$ component, and utilizing equations (\ref{con1}) and (\ref{con2}) gives a condition on $\mathcal{F}$ with respect to the lapse function
\begin{align}
\frac{\partial}{\partial r}\left(\frac{\mathcal{F}}{\alpha}\right)=0.
\end{align}
Integrating this gives $\alpha=g(t)\mathcal{F}$, where $g(t)$ is an arbitrary function of integration.  One can always utilize coordinate freedom to rescale the temporal coordinate such that $g(t)$ scales to unity.  Therefore, we find the relation $\mathcal{F}=\alpha$, implying the Killing vector is given by
\begin{align}
N^{\mu}=\frac{\partial}{\partial t}-\beta\frac{\partial}{\partial r}.
\end{align}
Finally, utilizing equation (\ref{con1}) along with the $\left(t,r\right)$ component of Killings equations gives
\begin{align}
\mathcal{L}_{n}\alpha=0.\label{con3}
\end{align}

Therefore, if the metric coefficients satisfy equations (\ref{con1}), (\ref{con2}) and (\ref{con3}), then the spacetime admits a hypersurface orthogonal, timelike Killing vector field, implying the spacetime is static.  We note that the standard method for searching for static solutions in a spacetime is to express the line element in a diagonal form, however we see the conditions given by the equations (\ref{con1}), (\ref{con2}) and (\ref{con3}) generally give $\beta\neq0$, implying the static spacetime is written in non-diagonal form.  One can further see this implies $\partial/\partial t$ is not the Killing vector, and in general there will be dependance on the temporal coordinate in the line element.  

Putting (\ref{con1}) and (\ref{con3}) into equation (\ref{LieE}) instantly gives the condition
\begin{align}
j=0.\label{jz}
\end{align}
That is, a static spherically symmetric fluid necessarily has vanishing ``heat flux''.  This makes physical sense as a flux intuitively implies a time rate of change, which is ruled out by the staticity condition.  

Now, equation (\ref{LieR}) reduces to
\begin{align}
\left(1+E\right)\left(\frac{\partial R}{\partial r}\right)^{2}=1-\frac{2M}{R},\label{done}
\end{align}
and furthermore from (\ref{LieM}) we find
\begin{align}
\mathcal{L}_{n}M=0,
\end{align}
where the mass function is now
\begin{align}
\frac{\partial M}{\partial r}=4\pi\rho\frac{\partial R}{\partial r}R^{2}.
\end{align}
Putting these results through equation (\ref{that}) and (\ref{Euler}) gives the generalization of the TOV equation for hydrostatic equilibrium  
\begin{align}
\frac{1}{\alpha}\frac{\partial\alpha}{\partial r}=&\frac{M+4\pi\left(P-2\Pi\right)R^{3}}{R^{2}\left(1-2M/R\right)}\,\,\frac{\partial R}{\partial r}\nonumber\\
=&\frac{-1}{\rho+P-2\Pi}\left[\frac{\partial}{\partial r}\left(P-2\Pi\right)-\frac{6\Pi}{R}\frac{\partial R}{\partial r}\right].\label{genTOV}
\end{align}
These equations are a generalization of those found in \cite{bowers74,poncedeleon87}, whereby we have extra freedom in the coordinates in form of the functions $\beta$ and $R$.  Finally, putting the staticity conditions through equation (\ref{cont}) gives
\begin{align}
\mathcal{L}_{n}\rho=0.\label{buch}
\end{align}

Summarily, the static system is given by equations (\ref{con1}), (\ref{con2}), (\ref{con3}) and (\ref{jz}-\ref{buch}).  

It is trivial to show if we let $R=r$, then equation (\ref{con1}) implies $\beta=0$ and the line element reduces to the standard diagonal form.  One can then see that $\alpha$, $E$, $\rho$ and $M$ all become functions of only the radial coordinate, which is consistent with $\partial/\partial t$ being the Killing vector.  Furthermore, the equations directly reduce to those given in \cite{bowers74,poncedeleon87}, who also provide some exact solutions of the equations by specifying various equations of state (also see \cite{dev02} and references therein).

\subsection{Exterior static region}
As with the non-static case, we can utilize the form of the interior to find the form of the exterior Schwarzschild region which matches continuously to the interior.  We begin by letting all the matter variables vanish at some finite radius, which implies $M=M_{s}$.  The left hand equation in (\ref{genTOV}) can then be integrated for the lapse function to give
\begin{align}
\alpha=\sqrt{1-\frac{2M_{s}}{R}},
\end{align}
where coordinate freedom has again been used to set a function of integration to unity.  It is interesting that there still exists coordinate freedom for the exterior, which would be determined when matching to a more specific form of the interior.  We choose here, as an illustrative example, to let $R=r$.  As mentioned, this implies $\beta=0$, and equations (\ref{con2}) and (\ref{done}) can then be used to show 
\begin{align}
E=-\frac{2M_{s}}{r}.
\end{align}
This implies the line element becomes the familiar Schwarzschild metric
\begin{align}
d\s^{2}=-\left(1-\frac{2M_{s}}{r}\right)dt^{2}+\frac{dr^{2}}{1-2M_{s}/r}+r^{2}d\Omega^{2}.
\end{align}
Therefore, a static interior general fluid, with $R=r$, matches continuously onto a Schwarzschild region in Schwarzschild coordinates.

\section{conclusion}
We have derived general equations governing a spherically symmetric system with an arbitrary matter distribution.  We showed that the system is underdetermined, and requires the input of more physics, most likely in the form of an extra conservation equation and thermodynamic considerations.  Furthermore, we have shown how the system reduces both to the Vaidya and Schwarzschild spacetimes by imposing only particular equations of state.  This implies entire collapse problems can be established utilizing the same coordinate patch throughout the spacetime.  This provides an alternative approach to the study of gravitational collapse of general fluids, which is usually achieved by establishing an interior metric that joins to the Schwarzschild or Vaidya metrics in the exterior using junction conditions at the interface (see for example \cite{bonnor89,herrera04,herrera06}).  However, this assumes one knows the form of the exterior metric {\it a priori}.  With the inclusion of more physics into the method derived herein, one will be able to determine the exterior region of the spacetime without prior knowledge.  

As an example of this, we describe the equations for the set-up of an initial value problem whereby on an initial spacelike slice a ball of fluid extends out to some finite radius, beyond which is vacuum extending to spacelike infinity.  As the system evolves, the interior is permitted to emit null radiation, and hence a region of Vaidya spacetime extends between the fluid and the vacuum.  While this scenario was described qualitatively, specific models of these scenarios should be made possible with numerical solutions of the equations derived herein.  This should allow for the modelling of more physical problems, whereby one does not require prior knowledge of the final solution in order to solve the problem.

This work opens avenues for further research into many currently unanswered questions.  In particular, the introduction of thermodynamics into the system will allow for the inclusion of conservation laws, such that knowledge of the exterior region is not required {\it a priori}.  Furthermore, being able to discuss the evolution of the temperature in all regions of the spacetime, including at $r=0$, at the apparent horizon, and also at the boundary of the various regions of the spacetime is an exciting prospect.  

Another aim of the current program of research is to extend our method beyond that of spherical symmetry.  In particular, we believe this method will be adaptable to the case of quasi-spherical symmetry, such that we can discuss interior regions for the Robinson-Trautman spacetime which represents outgoing gravitational radiation.  We expect the compactification diagram for the quasi-spherical case will be of the same form as figure \ref{diag}, however region III will represent the Robinson-Trautman solution and region I will represent some form of interior fluid with deviations from spherical symmetry.

Another future application of the current method is to introduce axisymmetry.  It is still an open question as to whether the Kerr solution represents the exterior to a rotating system.  Under the current paradigm, one would first derive general interior fluid equations, and simply let the matter terms tend to zero at some finite radius, thus determining the form of the exterior region of the spacetime.

\acknowledgments
The authors wish to thank R. Burston for useful discussions.  All calculations were checked using the computer algebra programme Maple.  



\bibliography{General_Fluid_Paper}

\end{document}